\documentclass[aps,prc, showpacs,amsmath,amssymb,nofootinbib]{revtex4} 
\usepackage{graphicx,latexsym,amssymb,epsfig}
\usepackage{color}
\usepackage{times}
\usepackage{diagbox} 
\usepackage[normalem]{ulem}
\usepackage{inputenc}
\usepackage{bm} 
\usepackage{makecell}
\usepackage{float}
\usepackage{url}
\usepackage{natbib}
\usepackage[colorlinks=true,citecolor=blue,urlcolor=blue,linkcolor=red]{hyperref}
\usepackage{amsmath}
\usepackage{lipsum}
\usepackage{graphicx}
\usepackage{dcolumn}
\usepackage{ulem} 
\usepackage{epstopdf}
\usepackage{epsfig}
\usepackage{float}
\usepackage{subfigure}
\usepackage{cancel}
\usepackage{multirow}
\usepackage[version=3]{mhchem}

\usepackage{soul}

\newcommand{\nc}{\newcommand}       
\nc{\vc}[1] {\mbox{\boldmath $#1$}} 
\nc{\del}       {\partial}              
\nc{\bra}       {\langle}               
\nc{\ket}       {\rangle}               
\nc{\bras}[1]   {\langle #1|}           
\nc{\kets}[1]   {|#1\rangle}            
\nc{\mapleft}[1]{           
	\smash{\mathop{\,          %
			\hbox to 1.5cm{\rightarrowfill}\, }\limits_{#1}}}
\nc{\beq}     {\begin{eqnarray}} \nc{\eeq}    {\end{eqnarray}}
\nc{\nn}      {\\\nonumber} \nc{\vs}      {\vspace{-0.275cm}}
\nc{\fra}    {\frac{1}{2}}
\nc{\mb}        {\mathbf}

\begin{document}
	
 \preprint{}
 
 \title{Investigations on the equation of state of neutron star  { matter} with density-dependent relativistic mean-field model}
 \author{Kaixuan Huang}
 \affiliation{School of Physics, Nankai University, Tianjin 300071,  China}
 
 \author{Jinniu Hu}~\email{hujinniu@nankai.edu.cn}
 \affiliation{School of Physics, Nankai University, Tianjin 300071,  China}
 \affiliation{Shenzhen Research Institute of Nankai University, Shenzhen 518083, China}

 \date{\today}
 
\begin{abstract}
       The neutron star as a supernova remnant is attracting high attention recently due to the gravitation wave detection and precise measurements about its mass and radius. In particular, the compact object with a mass of $2.50-2.67~M_\odot$ observed by LIGO Scientific and Virgo collaborations in GW190814, as well as the recent report of a light compact object with a mass and radius of $M=0.77^{+0.20}_{-0.17}M_{\odot}$ and $R=10.4^{+0.86}_{-0.78}$ km within the supernova remnant HESS J1731-347, have posed a great challenge to the investigations into the supranuclear matter. In the inner core region of the neutron star, the strangeness degrees of freedom, such as the hyperons, can be present, which is also named as a hyperonic star. In this work, the neutron star consisting of nucleons and leptons, and the hyperonic star including the hyperons will be studied in the framework of the density-dependent relativistic mean-field (DDRMF) model. Some popular DDRMF parameterizations will be adopted to investigate the properties of nuclear matter and the mass, radius, tidal deformability, and other properties of neutron star and hyperonic stars.
       We find that the maximum masses of neutron star calculated by DD-MEX, DD-MEX1, DD-MEX2, DD-MEXY and DD-LZ1 sets can be around $2.5-2.6~M_\odot$ with quite stiff  equations of state (EOSs) generated by their strong repulsive contributions from vector potentials at high densities. 
       Moreover, by investigating the influence of the crust EOS and core EOS on the neutron stars, we find that the observational data from HESS J1731-347 suggest the requirement of a crust EOS with a higher $L$ parameter and a core EOS with a lower $L$ parameter, and the $M-R$ relations from the constructed EOSs can also be consistent with the observables of PSR J0740+6620, PSR J0030+0451 from NICER and the GW170817 event.  
       With the inclusion of hyperons, the hyperonic star matter becomes softer compared to the neutron star matter. But the massive hyperonic star can also be obtained with DDRMF parameter sets if the vector coupling constants are strong.
\end{abstract} 

\pacs{21.10.Dr,  21.60.Jz,  21.80.+a}
 

\maketitle

\section{Introduction} 
With the rapid developments of astronomical technology, many significant achievements have been made in neutron star observation in the past decade. The massive neutron stars, whose masses are around $2M_\odot$, PSR J1614-2230 ($1.928\pm0.017M_\odot$)~\cite{demorest10}, PSR J0348+0432 ($2.01\pm0.04M_\odot$)~\cite{antoniadis13}, PSR J2215+5135 ($2.27^{+0.17}_{-0.15}M_\odot$)~\cite{linares18} and PSR J0740+6620 ($2.08\pm0.07M_\odot$)~\cite{cromartie20,miller21}, were measured within the relativistic Shapiro delay effect. The simultaneous measurements of mass-radius observations of the pulsars, PSR J0030+0451 \cite{miller19,riley19} and PSR J0740+6620 \cite{miller21,riley21}, using the pulse profile modeling of X-ray emissions from hot spots by NICER. What's more, the gravitational wave and the electromagnetic counterpart of a binary neutron-star merger were firstly detected by LIGO Scientific and Virgo Collaborations (LVC) as event GW170817~\cite{abbott17,abbott18,abbott19} and the dimensionless tidal deformability of $1.4M_\odot$ neutron star can be extracted from GW170817 as $190^{+390}_{-120}$~\cite{tews18,finstad18,fattoyev18}. In 2019, a new event of a compact binary merger with a $22.2-24.3 M_\odot$ black hole and a compact component with a mass of $2.50-2.67M_\odot$ was reported by LVC as GW190814~\cite{abbott20}. The secondary object of GW190814 attracts a lot of attentions, since it may be either the heaviest neutron star \cite{huang20} or the lightest black hole ever discovered~\cite{abbott20}. In addition to the massive neutron stars, there has been a growing focus on low-mass neutron stars. Recently, a central compact object (CCO) was reported within the supernova remnant HESS J1731-347. The estimated mass and radius of this object are $M=0.77^{+0.20}_{-0.17}M_{\odot}$ and $R=10.4^{+0.86}_{-0.78}$ km, respectively \cite{doroshenko22}.  There are various speculations regarding its internal constitution. Some studies have focused on investigating it as a quark star \cite{horvath23}, a hybrid star \cite{brodie23}, or a neutron star \cite{doroshenko22,huang24}.  

The structure of a static neutron star is separated into five regions. The outer layer is the atmosphere consisting of the atom and is very thin. In the next layer as the outer crust, the electrons of an atom are ionized and form the uniform Fermi gas, where the nuclei are immersed. With the density increasing, neutrons drip out from the neutron-rich nuclei and generate the neutron gas, which is called the inner crust. When the nucleon density approaches $\rho_0/2$, heavy nuclei dissolve and the neutron star matter becomes homogeneous, which corresponds to the outer core of neutron star. It plays an essential role in determining the mass and radius of the neutron star~\cite{grill12,bao15,pais15,yang19,ji21}. In the inner core region, the baryons including the strangeness degree of freedom, such as $\Lambda,~\Sigma$, and $\Xi$ hyperons, may appear, when the Fermi energies of nucleons are larger than their chemical potentials, which is also called as a hyperonic star.

The mass, radius, and tidal deformabilities of neutron stars and hyperonic stars can be determined by solving the Tolman–Oppenheimer–Volkoff (TOV) equation~\cite{tolman39,oppenheimer39} with the equation of state (EOS) of neutron star matter as input. Many attempts have been made to obtain the EOS of supranuclear matter in neutron stars from different models.  The relativistic mean-field (RMF) model has been widely successful in describing the properties of finite nuclei and naturally extends to high-density regimes with the $\rho$ meson, nonlinear terms of $\sigma$ and $\omega$ mesons, the coupling terms with $\rho$ meson to $\sigma$ or $\omega$ meson  introduced step by step \cite{walecka74,boguta77,serot79,sugahara94,horowitz01,meng06}. Furthermore, the nonlinear terms of various mesons could be replaced by the density-dependent meson-nucleon coupling constants in the density-dependent RMF (DDRMF), which consider the nuclear medium effect originated by the relativistic Brueckner-Hartree-Fock model~\cite{brockmann92}. 

The symmetry energy ($E_{\rm sym}$) and its density dependence play a crucial role in the EOS of neutron star matter because of its highly isospin-asymmetric nature. The behavior of $E_{\rm sym}$ near the saturation nucleon density ($\rho_0$) affects the structure of the neutron star crust \cite{oyamatsu07,bao14(89)} as well as the radius of neutron stars in the intermediate mass range \cite{lattimer01}. In addition to their impact on neutron stars, $E_{\rm sym}$ and its slope ($L$) can be constrained by terrestrial experiments. Recent measurements of the neutron skin thickness ($R_{\rm skin}$) of $^{208}$Pb by PREX-I and PREX-II resulted in $R_{\rm skin}^{208}=0.283\pm 0.071$ fm \cite{abrahamyan12, horowitz12, adhikari21}, leading to derived values of $E_{\rm sym}=38.1\pm4.7$ MeV and $L=106\pm37$ MeV based on the linear relation between $L$ and $R_{\rm skin}^{208}$ \cite{reed21,roca11,piekarewicz21}. Similarly, the CREX collaboration reported the neutron skin thickness of $^{48}$Ca as $R_{\rm skin}^{48}=0.121\pm 0.026$ fm \cite{adhikari22}, using the same method as PREX-II. Notably, the value of $L$ obtained from $^{48}$Ca is much smaller compared to that from PREX-II \cite{reinhard22,zhang23,lattimer23} and { the big difference between the two measurements brings significant challenges to the understanding of nuclear many-body methods}.

The hyperons were investigated from the 1980s by Glendenning in the framework of RMF model, where the coupling strengths between the mesons and baryons were simply generated by the quark power counting rules~\cite{glendenning82,glendenning85}. Actually, the appearances of hyperons are strongly dependent on the hyperon-nucleon and hyperon-hyperon potentials, which can be extracted from the properties of various hypernuclei. In the past 40 years, the RMF parameterizations about nucleon-nucleon and nucleon-hyperon interactions were largely improved through reproducing the ground-state properties of finite nuclei and above hypernuclei experimental data, that were adopted to investigate the properties of neutron stars and hyperonic  stars\cite{knorren95,schaffner96,shen02,dexheimer08,cavagnoli11,weissenborn12,providencia13,weissenborn13,fortin17,zhang18,li18,fortin20,shangguan21,tu21,lopes22}. With the discoveries of two-times-solar-mass neutron stars,  a "hyperon puzzle" was proposed, since the neutron star maximum mass will be reduced by about $20\%$ once the hyperons are self-consistently introduced in the nuclear many-body methods. Therefore, it is difficult to explain the existence of massive neutron stars with hyperons at the beginning and many schemes were raised to solve such a puzzle \cite{yamamoto14,logoteta15,logoteta19,zhang18,long12}. Recently, the role of vector meson including the strange quark, $\phi$ was also discussed in the hyperonic star with various hyperons \cite{biswal19,tu21}.

In this work, the neutron star and hyperonic star in the framework of DDRMF models will be systematically calculated with the most popular parameterizations, where the exchange mesons with strangeness quarks will be completely included and coupling strengths between the mesons and baryons will be constrained with the latest hypernuclei experimental data.  When calculating the properties of neutron stars, the EOS for light neutron stars will be constructed with DDVT model, incorporating the tensor coupling of vector mesons, to study the compact object in HESS J1731-347.  
This paper is arranged as follows. In Section \ref{sec2}, the formulas about the neutron star and hyperonic star with the DDRMF model are shown in detail. In Section \ref{sec3}, the parameterizations of the DDRMF model are listed and the properties of massive neutron stars, light neutron stars and hyperonic stars are calculated and discussed. In Section \ref{sec4}, the summary and conclusion will be given.

\section{The Density-dependent Relativistic Mean-field Model for Neutron Star}\label{sec2}
In the DDRMF model, the baryons interact with each other via exchanging various light mesons, including scalar-isoscalar meson ($\sigma$) with mass $m_{\sigma}$, vector-isoscalar meson ($\omega$)  with mass $m_{\omega}$, vector-isovector meson ($\rho$) with mass $m_{\rho}$, scalar-isoscalar meson ($\delta$) with mass $m_{\delta}$, and strange scalar and vector mesons ($\sigma^*$ and $\phi$) with mass $m_{\sigma^*}$ and $m_{\phi}$, respectively \cite{ring96,vretenar05,meng06,dutra14,lu11,liu18}. The baryons considered in the present calculation are nucleons ($n$ and $p$) and hyperons ($\Lambda,~\Sigma,~\Xi$). The Lagrangian density of DDRMF model is written as \cite{typel99,huang22}
\begin{equation}\label{eq.Ldd}
	\begin{aligned}
		\mathcal{L}
		=&\sum_{B}\overline{\psi}_{B}\left[\gamma^{\mu}\left(i\partial_{\mu}-\Gamma_{\omega B}(\rho_{B})\omega_{\mu}\right.\right.\\
		&\left.\left.-\Gamma_{\phi B}(\rho_{B})\phi_{\mu}-\frac{\Gamma_{\rho B}(\rho_{B})}{2}\vec{\rho}_{\mu}\vec{\tau}\right)\right. \\
		&\left.-\left(M_{B}-\Gamma_{\sigma B}(\rho_{B})\sigma-\Gamma_{\sigma^* B}(\rho_{B})\sigma^*-\Gamma_{\delta B}(\rho_{B})\vec{\delta}\vec{\tau}\right)\right]\psi_{B} \\
		&+\frac{1}{2}\left(\partial^{\mu}\sigma\partial_{\mu}\sigma-m_{\sigma}^2\sigma^2\right)
		+\frac{1}{2}\left(\partial^{\mu}\sigma^*\partial_{\mu}\sigma^*-m_{\sigma^*}^2{\sigma^*}^2\right) \\
		&+\frac{1}{2}\left(\partial^{\mu}\vec{\delta}\partial_{\mu}\vec{\delta}-m_{\delta}^2\vec{\delta}^2\right)
		-\frac{1}{4}W^{\mu\nu}W_{\mu\nu}+\frac{1}{2}m_{\omega}^2\omega_{\mu}\omega^{\mu} \\
		&-\frac{1}{4}\Phi^{\mu\nu}\Phi_{\mu\nu}+\frac{1}{2}m_{\phi}^2\phi_{\mu}\phi^{\mu}
		-\frac{1}{4}\vec{R}^{\mu\nu}\vec{R}_{\mu\nu}+\frac{1}{2}m_{\rho}^2\vec{\rho}_{\mu}\vec{\rho}^{\mu},
	\end{aligned}
\end{equation}
where $\psi_{B}$ represents the wave function of baryons {with mass $M_{B}$}. $\sigma,~\sigma^*,~\omega_{\mu},~\phi_{\mu}~\vec{\rho}_\mu$ denotes the fields of $\sigma,~\sigma^*,~\omega,~\phi$, and $\rho$ mesons, respectively. $W_{\mu\nu}$,  $\Phi_{\mu\nu}$, and $\vec{R}^{\mu\nu}$ are the anti-symmetry tensor fields of $\omega$, $\phi$ and $\rho$.
In nuclear matter, the tensor coupling between the vector meson and nucleon does not provide any contributions. Therefore, it is neglected in the present Lagrangian. The coupling constants of $\sigma$ and $\omega$ mesons are expressed as a fraction function of the total vector density, $\rho_{B}=\sum_{B}\rho^v_{B}$. In most of DDRMF parameterizations, such as DD2~\cite{typel10}, DD-ME1~\cite{niksic02}, DD-ME2~\cite{lalazissis05}, DD-MEX~\cite{taninah20}, PKDD~\cite{long04}, TW99~\cite{typel99}, and DDV, DDVT, DDVTD~\cite{typel20}, they are assumed as,
\begin{equation}\label{eq.sigcoup}
	\Gamma_{iN}(\rho_{B})=\Gamma_{iN}(\rho_{B0})f_i(x)
\end{equation}
with
\begin{equation}\label{eq.rhocoup}
	f_i(x)=a_i\frac{1+b_i(x+d_i)^2}{1+c_i(x+d_i)^2},~x=\rho_{B}/\rho_{B0},
\end{equation}
for $i=\sigma,~\omega$. $\rho_{B0}$ is the saturation density of symmetric nuclear matter.  The couplings $\Gamma_{iN}(\rho_{B0})$, and the coefficients $a_i,~b_i,~c_i$, and $d_i$ are obtained by fitting the properties of finite nuclei . Five constraints on the coupling constants $f_i(1)=1,~f_i^{''}(0)=0,~f_{\sigma}^{''}(1)=f_{\omega}^{''}(1)$ can reduce the numbers of independent parameters to three in Eq.~\eqref{eq.sigcoup}.  
For the isovector mesons $\rho$ and $\delta$, their density-dependent coupling constants are assumed to be,
\begin{gather} 
	\Gamma_{iN}(\rho_{B})=\Gamma_{iN}(\rho_{B0}){\rm exp}[-a_i(x-1)].
\end{gather}
While in other parameterizations, such as DD-LZ1~\cite{wei20}, the coefficient in front of fraction function, $\Gamma_i$ is fixed at $\rho_{B}=0$ for $i=\sigma,~\omega$:
\begin{gather} 
	\Gamma_{iN}(\rho_{B})=\Gamma_{iN}(0)f_i(x).
\end{gather}
There are only four constraint conditions as $f_i(0)=1$ and $f''_i(0)=0$ for $\sigma$ and $\omega$ coupling constants in DD-LZ1 set. The constraint $f''_{\sigma}(1)=f''_{\omega}(1)$ in previous parameter sets was removed in DD-LZ1 parameterization. For $\rho$ meson, its coupling constant is also changed accordingly as 
\begin{gather} 
	\Gamma_{\rho N}(\rho_{B})=\Gamma_{\rho N}(0){\rm exp}(-a_{\rho}x).
\end{gather}
To solve the nuclear many-body system in the DDRMF model, the mean-field approximation should be adopted, in which various mesons are treated as classical fields, 
$
\left\langle\sigma\right\rangle=\sigma,~\left\langle\sigma^*\right\rangle=\sigma^*,~\left\langle\omega_{\mu}\right\rangle=\omega,~\left\langle\phi_{\mu}\right\rangle=\phi,
~\left\langle\vec{\rho}_{\mu}\right\rangle=\rho,~\langle\delta\rangle=\delta.
$
The space components of the vector mesons are removed in the parity conservation system. In addition, the spatial derivatives of baryons and mesons are neglected in the infinite nuclear matter due to the transformation invariance. Finally, with the Euler-Lagrange equation, the equations of motion for baryons and mesons are obtained,
\begin{align}\label{eq.EOM}
	&\left[i\gamma^{\mu}\partial_{\mu}-\gamma^0\left(\Gamma_{\omega B}(\rho_{B})\omega+\Gamma_{\phi B}(\rho_{B})\phi\right.\right.\nonumber\\
	&\left.\left.+\frac{\Gamma_{\rho B}(\rho_{B})}{2}\rho\tau_3+\Sigma_R(\rho_{B})\right)-M_{B}^{*}\right]\psi_{B}=0.\nonumber\\
	&m_{\sigma}^2\sigma=\sum_{B}\Gamma_{\sigma B}(\rho_{B})\rho^s_{B},\nonumber\\
	&m_{\sigma^*}^2\sigma^*=\sum_{B}\Gamma_{\sigma^* B}(\rho_{B})\rho^s_{B},\nonumber\\
	&m_{\omega}^2\omega=\sum_{B}\Gamma_{\omega B}(\rho_{B})\rho_{B},\nonumber\\
	&m_{\phi}^2\phi=\sum_{B}\Gamma_{\phi B}(\rho_{B})\rho_{B},\nonumber\\
	&m_{\rho}^2\rho=\sum_{B}\frac{\Gamma_{\rho B}(\rho_{B})}{2}\rho^{3}_{B},\nonumber\\
	&m_{\delta}^2\delta=\sum_{B}\Gamma_{\delta B}(\rho_{B})\rho^{s3}_{B}.
\end{align}
The rearrangement contribution, $\Sigma_R$, is expressed as
\begin{align}
	\Sigma_R(\rho_{B})=&-\frac{\partial\Gamma_{\sigma B}(\rho_{B})}{\partial\rho_{B}}\sigma\rho^s_{B}
	-\frac{\partial\Gamma_{\sigma^* B}(\rho_{B})}{\partial\rho_{B}}\sigma^*\rho^s_{B}\nonumber\\
	&-\frac{\partial\Gamma_{\delta B}(\rho_{B})}{\partial\rho_{B}}\delta\rho^{s3}_{B}
	+\frac{1}{2}\frac{\partial\Gamma_{\rho B}(\rho_{B})}{\partial\rho_{B}}\rho\rho^{3}_{B}\nonumber\\
	&+\left[\frac{\partial\Gamma_{\omega B}(\rho_{B})}{\partial\rho_{B}}\omega+\frac{\partial\Gamma_{\phi B}(\rho_{B})}{\partial\rho_{B}}\phi\right]\rho_{B},
\end{align}
where the scalar, vector densities, and their isospin components are
\begin{align} 
	\rho^s_{B}=\left\langle\overline{\psi}_{B}\psi_{B}\right\rangle,~~~~
	&\rho^{s3}_{B}=\left\langle\overline{\psi}_{B}\tau_3\psi_{B}\right\rangle,\nonumber\\
	\rho_{B}=\left\langle\psi^{\dag}_{B}\psi_{B}\right\rangle,~~~~
	&\rho^{3}_{B}=\left\langle\psi^{\dag}_{B}\tau_3\psi_{B}\right\rangle.
\end{align} 
$\tau_3$ in the above equation is the isospin third component of the baryon species $B$.  
The effective masses of baryons are dependent on the scalar mesons $\sigma$, $\sigma^*$ and $\delta$,
\begin{align}
	M_{B}^{*}&=M_{B}-\Gamma_{\sigma B}(\rho_{B})\sigma-\Gamma_{\sigma^* B}(\rho_{B})\sigma^*-\Gamma_{\delta B}(\rho_{B})\delta\tau_3.
\end{align}
Because of the mass-energy relation, the corresponding effective energies of baryons are   
\begin{gather}\label{eq.effene}
	E_{FB}^{*}=\sqrt{k_{FB}^2+(M_{B}^{*})^2},
\end{gather}
where $k_{FB}$ is the Fermi momentum of baryons. 

With the energy-momentum tensor in a uniform system, the energy density, $\mathcal{E}$ and pressure, $P$ of nuclear matter in DDRMF model are obtained respectively as
\begin{align}
	\mathcal{E}=&\frac{1}{2}m_{\sigma}^2\sigma^2+\frac{1}{2}m_{\sigma^*}^2{\sigma^*}^2
	-\frac{1}{2}m_{\omega}^2\omega^2-\frac{1}{2}m_{\phi}^2\phi^2\nonumber\\
	&-\frac{1}{2}m_{\rho}^2\rho^2+\frac{1}{2}m_{\delta}^2\delta^2+\Gamma_{\omega B}(\rho_{B})\omega\rho_{B}\nonumber\\
	&+\Gamma_{\phi B}(\rho_{B})\phi\rho_{B} +\frac{\Gamma_{\rho}(\rho_{B})}{2}\rho\rho_3+\mathcal{E}_{\rm kin}^{B},\nonumber\\
	P=
	&\rho_{B}\Sigma_{R}(\rho_{B})-\frac{1}{2}m_{\sigma}^2\sigma^2-\frac{1}{2}m_{\sigma^*}^2{\sigma^*}^2\nonumber\\
	&+\frac{1}{2}m_{\omega}^2\omega^2+\frac{1}{2}m_{\phi}^2\phi^2+\frac{1}{2}m_{\rho}^2\rho^2\nonumber\\
	&-\frac{1}{2}m_{\delta}^2\delta^2+P_{\rm kin}^{B},
\end{align}
where the contributions from kinetic energy are 
\begin{align}
	\mathcal{E}_{\rm kin}^{B}&=\frac{\gamma}{2\pi^2}\int_{0}^{k_{FB}}k^2\sqrt{k^2+{M_{B}^{*}}^{2}}dk\nonumber\\
	&=\frac{\gamma}{16\pi^2}\left[k_{FBi}E_{FB}^{*}\left(2k_{FB}^2+{M_{B}^{*}}^2\right)+{M_{B}^{*}}^4{\rm ln}\frac{M_{B}^{*}}{k_{FB}+E_{FB}^{*}}\right], \\
	P_{\rm kin}^{B}&=\frac{\gamma}{6\pi^2}\int_{0}^{k_{FB}}\frac{k^4 dk}{\sqrt{k^2+{M_{B}^{*}}^{2}}}\nonumber\\
	&=\frac{\gamma}{48\pi^2}\left[k_{FB}\left(2k_{FB}^2-3{M_{B}^{*}}^2\right)E_{FB}^{*}+3{M_{B}^{*}}^{4}{\rm ln}\frac{k_{FB}+E_{FB}^{*}}{M_{B}^{*}}\right].
\end{align}
$\gamma=2$ is the spin degeneracy factor. 

The binding energy per nucleon for the symmetric nuclear matter can be defined by
\begin{align}
	\frac{E}{A}&=\frac{\mathcal{E}}{\rho_{B}}-M.
\end{align}
The symmetry energy is calculated by 
\begin{equation}\label{eq.esym}
	E_{\rm sym}=\frac{1}{2}\frac{\partial^2E/A}{\partial\alpha^2}
	=\frac{k_F^2}{6E_F^*}+\frac{\Gamma_{\rho N}^2}{8m_{\rho}^2}\rho_N,
\end{equation}
where $\alpha$ is the asymmetry factor, defined as $\alpha=(\rho_{Bn}-\rho_{Bp})/(\rho_{Bn}+\rho_{Bp})$ and its slope at the saturation point, $L$ is given by
\begin{gather}
	L=3\rho_{B}\frac{\partial E_{\rm sym}}{\partial\rho_B}\Bigg|_{\rho_B=\rho_{B0}}.
\end{gather} 

In the uniform neutron star matter, the compositions of baryons and leptons are determined by the requirements of charge neutrality and $\beta$-equilibrium conditions. All baryon octets $(n,~p,~\Lambda,~\Sigma^-,~\Sigma^0,~\Sigma^+,~\Xi^-,~\Xi^0)$ and leptons ($e,~\mu$) will be included in this work. The $\beta$-equilibrium conditions can be expressed by~\cite{glendenning85,shen02}
\begin{gather} \label{eq.betacond}
	\mu_p=\mu_{\Sigma^+}=\mu_n-\mu_e,\nonumber\\
	\mu_{\Lambda}=\mu_{\Sigma^0}=\mu_{\Xi^0}=\mu_n,\nonumber\\
	\mu_{\Sigma^-}=\mu_{\Xi^-}=\mu_n+\mu_e,\nonumber\\
	\mu_{\mu}=\mu_e,	
\end{gather} 
where $\mu_i$ can be derived from the thermodynamics equations at zero temperature,  
\begin{align}\label{eq.chem}
	\mu_{B}&=\sqrt{k_{FB}^2+M_{B}^{*2}}+\left[\Gamma_{\omega B}(\rho_{B})\omega+\Gamma_{\phi B}(\rho_{B})\phi\right. \nonumber\\
	&\left.+\frac{\Gamma_{\rho B}(\rho_{B})}{2}\rho^3_B+\Sigma_R(\rho_{B})\right],\nonumber\\ 
	\mu_l&=\sqrt{k_{Fl}^2+m_l^{2}}.
\end{align} 
The charge neutrality condition has the following form,
\begin{gather}\label{eq.chargecond}
	\rho_{p}+\rho_{\Sigma^+}=\rho_e+\rho_{\mu}+\rho_{\Sigma^-}+\rho_{\Xi^-}.
\end{gather}

The total energy density and pressure of neutron star matter will be obtained as a function of baryon density within the constraints of Eqs.~\eqref{eq.betacond} and \eqref{eq.chargecond}. The Tolman-Oppenheimer-Volkoff (TOV) equation describes a spherically symmetric star in the gravitational equilibrium from general relativity~\cite{oppenheimer39,tolman39},
\begin{gather}\label{eq.tov}
	\frac{dP}{dr}=-\frac{GM(r)\mathcal{E}(r)}{r^2}\frac{\left[1+\frac{P(r)}{\mathcal{E}(r)}\right]\left[1+\frac{4\pi r^3P(r)}{M(r)}\right]}{1-\frac{2GM(r)}{r}},\nonumber\\
	\frac{dM(r)}{dr}=4\pi r^2\mathcal{E}(r),
\end{gather}
where $P$ and $M$ are the pressure and mass of a neutron star at the position $r$ {and $G$ is the gravitational
	constant}. Furthermore, the tidal deformability becomes a typical property of a neutron star after the observation of the gravitational wave from a binary neutron-star (BNS) merger, which characterizes the deformation of a compact object in an external gravity field generated by another star. The tidal deformability of a neutron star is reduced as a dimensionless form~\cite{hinderer08,postnikov10},
\begin{gather} 
	\Lambda=\frac{2}{3}k_2C^{-5}.
\end{gather}
where $C=GM/R$ is the compactness parameter. The second order Love number $k_2$ is given by
\begin{align} 
	k_2=&\frac{8C^5}{5}(1-2C)^2\left[2+2C(\mathcal{Y}_R-1)-\mathcal{Y}_R\right]\nonumber\\
	&\Big\{2C\left[6-3\mathcal{Y}_R+3C(5\mathcal{Y}_R-8)\right]\nonumber\\
	&+4C^3\left[13-11\mathcal{Y}_R+C(3\mathcal{Y}_R-2)+2C^2(1+\mathcal{Y}_R)\right]\nonumber\\
	&+3(1-2C)^2\left[2-\mathcal{Y}_R+2C(\mathcal{Y}_R-1){\rm ln}(1-2C)\right]\Big\}^{-1}.
\end{align}
Here, $\mathcal{Y}_R=y(R)$. $y(r)$ satisfies the following first-order differential equation,
\begin{equation}
	r\frac{d y(r)}{dr} + y^2(r)+y(r)F(r) + r^2Q(r)=0,
\end{equation}
where $F(r)$ and $Q(r)$ are functions related to the pressure and energy density
\begin{align}
	F(r) & = \left[1-\frac{2M(r)}{r}\right]^{-1} 
	\left\{1-4\pi r^2[\mathcal{E}(r)-P(r)]\right\} ,\\ 
	\nonumber 
	r^2Q(r) & =  \left\{4\pi r^2 \left[5\mathcal{E}(r)+9P(r)+\frac{\mathcal{E}(r)
		+P(r)}{\frac{\partial P}{\partial \mathcal{E}}(r)}\right]-6\right\}\\ 
	\nonumber 
	&\times \left[1-\frac{2M(r)}{r}\right]^{-1} -\left[\frac{2M(r)}{r} +2\times4\pi r^2 P(r) \right]^2\\ 
	\nonumber 
	&\times \left[1-\frac{2M(r)}{r}\right]^{-2} .
\end{align}
The second Love number corresponds to the initial condition $y(0)=2$. It is also related to the speed of sound in compact matter, $c_s$,
\begin{gather}\label{vs2}
	c_s^2=\frac{\partial P(\varepsilon)}{\partial{\mathcal{E}}}.
\end{gather}

\section{Results and Discussions}\label{sec3}
\subsection{The DDRMF Parameterizations} 
\begin{table}[htbp] 
\centering
\caption{ Masses of nucleons and mesons, meson coupling constants, and the nuclear saturation densities in various DDRMF models. The coefficients of meson coupling constants, $\Gamma_i$ in DD-LZ1 are the values at zero density, while other parameter sets are dependent on the values at nuclear saturation densities.}\label{table.para}
\scalebox{1}{
	\rotatebox{90}{
	\begin{tabular}{lcccccccccccccccccc}
		\hline\hline
		&DD-LZ1     &             &DD-MEX     &DD-MEX1    &DD-MEX2    &DD-MEXY     &DD-ME2   &DD-ME1   &DD2   &PKDD    &TW99       &DDV     &DDVT      &DDVTD        \\
		\hline
		$m_n[\rm MeV]$        &938.900000 &$m_n$        &939.0000   &939.000000 &939.000000 &939.000000 &939.0000  &939.0000  &939.56536   &939.5731  &939.0000  &939.565413  &939.565413  &939.565413 \\
		$m_p[\rm MeV]$        &938.900000 &$m_p$        &939.0000   &939.000000 &939.000000 &939.000000 &939.0000  &939.0000  &938.27203   &938.2796  &939.0000  &938.272081  &938.272081  &938.272081 \\
		$m_{\sigma}[\rm MeV]$ &538.619216 &$m_{\sigma}$ &547.3327   &553.714785 &551.087886 &551.321796 &550.1238  &549.5255  &546.212459  &555.5112  &550.0000  &537.600098  &502.598602  &502.619843 \\
		$m_{\omega}[\rm MeV]$ &783.0000   &$m_{\omega}$ &783.0000   &783.000000 &783.000000 &783.000000 &783.0000  &783.0000  &783.0000    &783.0000  &783.0000  &783.0000    &783.0000    &783.0000   \\
		$m_{\rho}[\rm MeV]$   &769.0000   &$m_{\rho}$   &763.0000   &763.000000 &763.000000 &763.000000 &763.0000  &763.0000  &763.0000    &763.0000  &763.0000  &763.0000    &763.0000    &763.0000   \\
		$m_{\delta}[\rm MeV]$ &---        &$m_{\delta}$ &---        &         &         &          &---       &---       &---       &---       &---       &---         &---         &980.0000   \\
		$\Gamma_{\sigma}(0)$  &12.001429  &$\Gamma_{\sigma}(\rho_{{B}0})$ &10.7067  &10.668226 &10.476976 &10.411867 &10.5396 &10.4434 &10.686681 &10.7385 &10.72854 &10.136960  &8.382863  &8.379269  \\
		$\Gamma_{\omega}(0)$  &14.292525  &$\Gamma_{\omega}(\rho_{{B}0})$ &13.3388  &13.107751 &12.903532 &12.803298 &13.0189 &12.8939 &13.342362 &13.1476 &13.29015 &12.770450  &10.987106 &10.980433 \\
		$\Gamma_{\rho}(0)$    &15.150934  &$\Gamma_{\rho}(\rho_{{B}0})$   &7.2380   &7.283016  &8.201438  &7.38434   &7.3672  &7.6106  &7.25388   &8.5996  &7.32196  &7.84833    &7.697112  &8.06038 \\
		$\Gamma_{\delta}(0)$  &---        &$\Gamma_{\delta}(\rho_{{B}0})$ &---      &---  &---  &---   &---      &--       &---        &---      &---       &---        &---        &0.8487420   \\
		\hline                                                                            
		$\rho_{{B}0}[\rm fm^{-3}]$ &0.1581  &$\rho_{{B}0}$            &0.1520   &0.1510  &0.1500 &0.1500   &0.1520  &0.1520  &0.1490   &0.1496  &0.1530  &0.1511  &0.1536  &0.153  \\
		\hline                                                                                    
		$a_{\sigma}$          &1.062748   &$a_{\sigma}$  &1.3970  &1.392047  &1.408690  &1.437200  &1.3881  &1.3854  &1.357630   &1.327423  &1.365469  &1.20993     &1.20397     &1.19643     \\
		$b_{\sigma}$          &1.763627   &$b_{\sigma}$  &1.3350  &2.107233  &1.506460  &2.059712  &1.0943  &0.9781  &0.634442   &0.435126  &0.226061  &0.21286844  &0.19210314  &0.19171263  \\
		$c_{\sigma}$          &2.308928   &$c_{\sigma}$  &2.0671  &3.156692  &2.337477  &3.210289  &1.7057  &1.5342  &1.005358   &0.691666  &0.409704  &0.30798197  &0.27773566  &0.27376859  \\
		$d_{\sigma}$          &0.379957   &$d_{\sigma}$  &0.4016  &0.324955  &0.377629  &0.322231  &0.4421  &0.4661  &0.575810   &0.694210  &0.901995  &1.04034342  &1.09552817  &1.10343705  \\
		$a_{\omega}$          &1.059181   &$a_{\omega}$  &1.3936  &1.382434  &1.404071  &1.431375  &1.3892  &1.3879  &1.369718   &1.342170  &1.402488  &1.23746     &1.16084     &1.16693     \\
		$b_{\omega}$          &0.418273   &$b_{\omega}$  &1.0191  &1.880412  &1.349038  &1.943724  &0.9240  &0.8525  &0.496475   &0.371167  &0.172577  &0.03911422  &0.04459850  &0.02640016  \\
		$c_{\omega}$          &0.538663   &$c_{\omega}$  &1.6060  &2.811153  &2.100795  &3.025356  &1.4620  &1.3566  &0.817753   &0.611397  &0.344293  &0.07239939  &0.06721759  &0.04233010  \\
		$d_{\omega}$          &0.786649   &$d_{\omega}$  &0.4556  &0.344348  &0.398334  &0.331934  &0.4775  &0.4957  &0.638452   &0.738376  &0.983955  &2.14571442  &2.22688558  &2.80617483  \\
		$a_{\rho}$            &0.776095   &$a_{\rho}$    &0.6202  &0.561222  &0.193540  &0.532267  &0.5647  &0.5008  &0.518903   &0.183305  &0.5150    &0.35265899  &0.54870200  &0.55795902  \\
		$a_{\delta}$          &---        &$a_{\delta}$  &---    &---   &---   &---   &---       &---        &---          &---         &---         &---         &---         &0.55795902    \\
		\hline\hline
	\end{tabular}}}
\end{table}
We list some of the DDRMF parameterizations in Table~\ref{table.para}, where the DD2~\cite{typel10}, DD-ME1~\cite{niksic02}, DD-ME2~\cite{lalazissis05}, DD-MEX~\cite{taninah20}, PKDD~\cite{long04}, TW99~\cite{typel99}, DDV, DDVT, DDVTD~\cite{typel20}, and DD-LZ1~\cite{wei20} have been applied to study the properties of nuclear matter and the neutron stars in our previous works \cite{huang20,huang22}. DD-MEX1, DD-MEX2, and DD-MEXY ~\cite{taninah23} are the new DDRMF parameter sets based on the DD-MEX set. In DDVT and DDVTD sets, the tensor coupling between the vector meson and nucleon was included. The scalar-isovector meson, $\delta$ was taken into account in DDVTD set. 

\begin{figure*}[htbp!]
	\centering
	\includegraphics[scale=0.45]{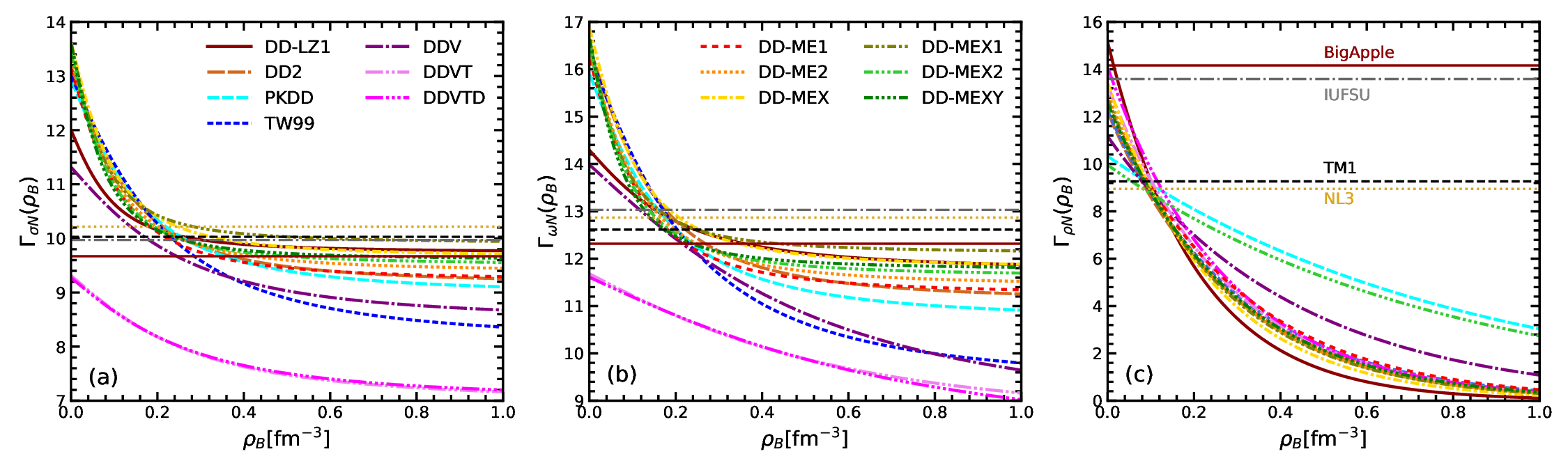}
	\caption{The coupling constants of $\omega,~\sigma$, and $\rho$ mesons as functions of vector density in various DDRMF models and several nonlinear RMF models.}\label{fig.1}
\end{figure*}

The density-dependent behaviors of the coupling constants as functions of the vector density are shown in Fig.~\ref{fig.1}. It can be found that all of these coupling constants decrease when the nuclear density becomes larger due to the nuclear medium effect. For the $\rho$ meson coupling constants in panel (c), all parameter sets have very similar density-dependent behaviors in the whole density region. In DDVT and DDVTD, the tensor coupling constants play obvious roles in finite nuclei due to their derivative forms, while they do not provide any contribution in nuclear matter. Their coupling constants of $\sigma$ and $\omega$ mesons in panel (a) and panel (b) are dramatically smaller than other sets. Furthermore, the coupling constants from several typical nonlinear RMF models, NL3~\cite{lalazissis97}, TM1~\cite{sugahara94}, IUFSU~\cite{fattoyev10}, and BigApple~\cite{fattoyev20} are also shown to compare their differences with those in DDRMF model. At low density region, the coupling constants in DDRMF models are usually stronger than those in nonlinear RMF modes, while weaker at higher density.  

\begin{table*}[htbp]
	\small
	\centering
	\caption{Nuclear matter properties at saturation density generated by DDRMF parameterizations.}\label{table.sat}
	\scalebox{1}{
		\begin{tabular}{lcccccccccccccccc}
			\hline\hline
			&$\rho_{{\rm B}0}[\rm fm^{-3}]$ ~~&$E/A[\rm MeV]$   ~~&$K_0[\rm MeV]$  ~~&$E_{\rm sym}[\rm MeV]$  ~~&$L_0[\rm MeV]$ ~~&$M_n^{*}/M$ ~~&$M_p^{*}/M$\\
			\hline 
			DD-LZ1       &0.1581  ~~&-16.0598  ~~&231.1030  ~~&31.3806  ~~&42.4660  ~~&0.5581  ~~&0.5581  \\
			DD-MEX       &0.1519  ~~&-16.0973  ~~&267.3819  ~~&32.2238  ~~&46.6998  ~~&0.5554  ~~&0.5554  \\
			DD-MEX1      &0.1505  ~~&-16.0368  ~~&291.1968  ~~&31.8312  ~~&53.4254  ~~&0.5709  ~~&0.5709   \\
			DD-MEX2      &0.1520  ~~&-16.0376  ~~&255.0925  ~~&35.2921  ~~&86.8244  ~~&0.5780  ~~&0.5780   \\
			DD-MEXY      &0.1535  ~~&16.0243   ~~&367.9365  ~~&32.0355  ~~&53.2101  ~~&0.5811  ~~&0.5811   \\
			DD-ME2       &0.1520  ~~&-16.1418  ~~&251.3062  ~~&32.3094  ~~&51.2653  ~~&0.5718  ~~&0.5718  \\
			DD-ME1       &0.1522  ~~&-16.2328  ~~&245.6657  ~~&33.0899  ~~&55.4634  ~~&0.5776  ~~&0.5776  \\
			DD2          &0.1491  ~~&-16.6679  ~~&242.8509  ~~&31.6504  ~~&54.9529  ~~&0.5627  ~~&0.5614  \\
			PKDD         &0.1495  ~~&-16.9145  ~~&261.7912  ~~&36.7605  ~~&90.1204  ~~&0.5713  ~~&0.5699  \\
			TW99         &0.1530  ~~&-16.2472  ~~&240.2022  ~~&32.7651  ~~&55.3095  ~~&0.5549  ~~&0.5549  \\
			DDV          &0.1511  ~~&-16.9279  ~~&239.9522  ~~&33.5969  ~~&69.6813  ~~&0.5869  ~~&0.5852  \\
			DDVT         &0.1536  ~~&-16.9155  ~~&239.9989  ~~&31.5585  ~~&42.3414  ~~&0.6670  ~~&0.6657  \\
			DDVTD        &0.1536  ~~&-16.9165  ~~&239.9137  ~~&31.8168  ~~&42.5829  ~~&0.6673  ~~&0.6660  \\
			\hline
			BSk19        &0.1596 ~~&-16.08    ~~&237.3  ~~&30.0     ~~&31.9     ~~&0.8    ~~&0.8   \\         
			BSk20        &0.1596 ~~&-16.08    ~~&241.4     ~~&30.0     ~~&34.7     ~~&0.8     ~~&0.8   \\ 
			BSk21        &0.1582 ~~&-13.05    ~~&245.8     ~~&30.0     ~~&46.6     ~~&0.8     ~~&0.8   \\ 
			\hline
			IUFSU        &0.1545  ~~&-16.3973 ~~&230.7491  ~~&31.2851  ~~&47.1651   ~~&0.6095  ~~&0.6095 \\ 
			\hline\hline    
	\end{tabular}}
\end{table*}
The saturation properties of symmetric nuclear matter calculated with different DDRMF effective interactions are listed in Table~\ref{table.sat}, i.e. the saturation density, $\rho_0$, the binding energy per nucleon, $E/A$, incompressibility, $K_0$, symmetry energy, $E_\text{sym}$, the slope of symmetry energy, $L$, and the effective neutron and proton masses, $M^*_n$ and $M^*_p$. 
Notably, DD-MEX2 and PKDD exhibit a distinct $L$ compared to other DDRMF sets, while DDVT demonstrates a notable difference in effective nucleon mass ($M^*_N$). The fitting process of DD-MEX2 did not account for constraints from pure neutron matter and neutron skin, whereas the tensor coupling terms in DDVT suppress the magnitude of the $\sigma$ field, resulting in a larger effective nucleon mass. To examine the influence of effective mass on neutron star properties, we also considered three non-relativistic density-functional theory parameterizations, namely BSk19, BSk20, and BSk21 \cite{goriely10,potekhin13}, based on Skyrme-type effective interactions. These parameterizations have an relatively large effective mass of $0.8M_N$ at saturation density and relatively small $L$ values of $31.90$, $34.70$, and $46.60$ MeV, respectively. 

\begin{figure*}[htbp!]
	\centering
	\includegraphics[scale=0.5]{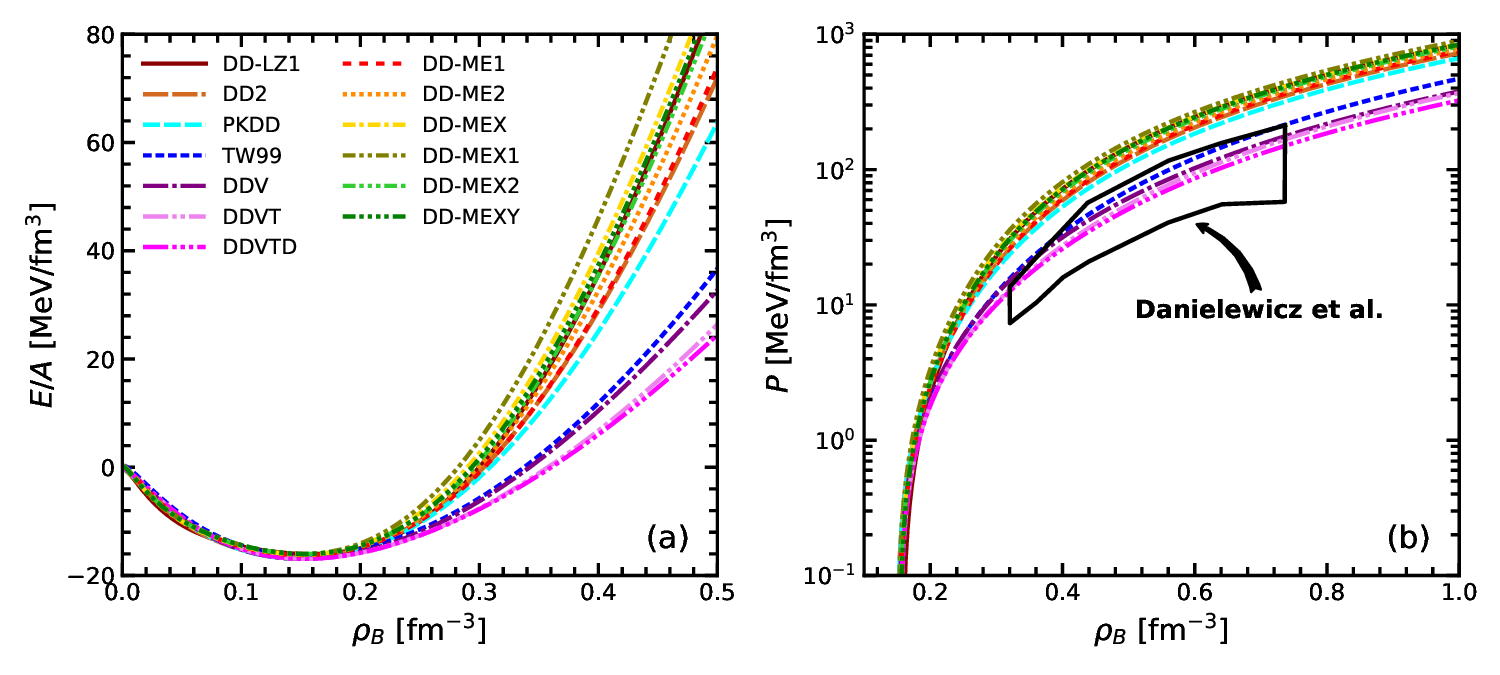}
	\caption{The binding energies per nucleon the pressures as functions of vector density for symmetric nuclear matter with various DDRMF models.}\label{fig.2}
\end{figure*}

The binding energies per nucleon as functions of vector density for symmetric nuclear matter are plotted in panel (a) of Fig~.\ref{fig.2} with the present DDRMF parameterizations. These EOSs of nuclear matter below $0.2$ fm$^{-3}$ are almost identical since all the parameters were determined by properties of finite nuclei, whose central density is around nuclear saturation density $\rho_{B0}\sim0.15$ fm$^{-3}$. Their differences increase from $0.30$ fm$^{-3}$ and they are separated into the softer group with DDV, DDVT, DDVTD { and TW99}, and the stiffer group with DD2, DD-ME1, DD-ME2, DD-MEX, DD-MEX1, DD-MEX2, DD-MEXY and DD-LZ1 since the scalar and vector coupling strengths in softer group sets are obviously weaker than those in stiffer group sets \cite{huang20}. 

In panel (b) of Fig.~\ref{fig.2}, the pressures in nuclear matter as functions of density are shown and compared to the constraints from heavy-ion collisions at $2-4\rho_{B0}$ by Danielewicz et al.~\cite{danielewicz02}. We can find that the EOSs from the softer group sets are completely consistent with the experiment data, while the other group is indeed stiffer than the heavy-ion collisions constraints.  However, we want to emphasize here that the constraints from the heavy-ion collisions are strongly model-dependent, which is determined by many inputs, such as the $NN$ interaction. To our knowledge, there were few investigations about heavy-ion collisions, which adopted the RMF model as $NN$ interaction. { It} cannot certainly claim that the EOSs generated by DD2, DD-ME1, DD-ME2, DD-MEX, DD-MEX1, DD-MEX2, DD-MEXY and DD-LZ1 parameterizations are clearly excluded by the constraints of heavy-ion collisions. 

For the hyperonic star matter with strangeness degree of freedom, the hyperon masses are chosen as $m_{\Lambda}=1115.68$ MeV, $m_{\Sigma^+}=1189.37$ MeV, $m_{\Sigma^0}=1192.64$ MeV, $m_{\Sigma^-}=1197.45$ MeV, $m_{\Xi^0}=1314.86$ MeV, and  $m_{\Xi^-}=1321.71$ MeV~\cite{zyla20}, while the masses of strange mesons, $\phi$ and $\sigma^*$ are taken as $m_{\phi}=1020$ MeV and $m_{\phi}=980$ MeV. We adopt the values from the SU(6) symmetry for the coupling constants between hyperons and vector mesons~\cite{dover84} both in nonlinear RMF and  DDRMF models. Here, the coupling constants of the DDRMF model are taken as an example,
\begin{gather}
	\Gamma_{\omega\Lambda}=\Gamma_{\omega\Sigma}=2\Gamma_{\omega\Xi}=\frac{2}{3}\Gamma_{\omega N},\nonumber\\
	2\Gamma_{\phi\Sigma}=\Gamma_{\phi\Xi}=-\frac{2\sqrt{2}}{3}\Gamma_{\omega N},~\Gamma_{\phi N}=0,\nonumber\\
	\Gamma_{\rho\Lambda}=0,~\Gamma_{\rho\Sigma}=2\Gamma_{\rho\Xi}=2\Gamma_{\rho N},\nonumber\\
	\Gamma_{\delta\Lambda}=0,~\Gamma_{\delta\Sigma}=2\Gamma_{\delta\Xi}=2\Gamma_{\delta N},
\end{gather}  
where $\Gamma_{iN}$ has been defined in Eq. \eqref{eq.sigcoup}-Eq. \eqref{eq.rhocoup} for DDRMF models. The coupling constants of hyperon and scalar mesons are constrained by the hyperon-nucleon potentials in symmetric nuclear matter, $U_Y^N$, which are defined by \cite{huang22}
\begin{equation}\label{pot}
	U_Y^N(\rho_{{B}0})=-R_{\sigma Y}\Gamma_{\sigma N}(\rho_{{B}0})\sigma_0+R_{\omega Y}\Gamma_{\omega N}(\rho_{{B}0})\omega_0,
\end{equation}
where $\Gamma_{\sigma N}(\rho_{{B}0}),~\Gamma_{\omega N}(\rho_{{B}0}),~\sigma_0,~\omega_0$ are the values of coupling strengths and $\sigma,~\omega$ meson fields at the saturation density. $R_{\sigma Y}$ and $R_{\omega Y}$ are defined as $R_{\sigma Y}=\Gamma_{\sigma Y}/\Gamma_{\sigma N}$ and $R_{\omega Y}=\Gamma_{\omega Y}/\Gamma_{\omega N}$. We choose the hyperon-nucleon potentials of $\Lambda,~\Sigma$ and $\Xi$ as $U_{\Lambda}^N=-30$ MeV, $U_{\Sigma}^N=+30$ MeV and $U_{\Xi}^N=-14$ MeV, respectively from the recent hypernuclei experimental observables~\cite{fortin17,hu22,khaustov00}. 

The coupling constants between $\Lambda$ and $\sigma^*$ is generated by the value of the $\Lambda\Lambda$ potential in pure $\Lambda$ matter, $U_{\Lambda}^{\Lambda}$ at nuclear saturation density, which is given as
\begin{align}
	U_{\Lambda}^{\Lambda}(\rho_{{B}0})=&-R_{\sigma\Lambda}\Gamma_{\sigma N}(\rho_{{B}0})\sigma_0-R_{\sigma^*\Lambda}\Gamma_{\sigma N}(\rho_{{B}0}){\sigma^*_0}\nonumber\\
	&+R_{\omega Y}\Gamma_{\omega N}(\rho_{{B}0})\omega_0+R_{\phi\Lambda}\Gamma_{\omega N}(\rho_{{B}0})\phi_0,
\end{align}
We similarly define that $R_{\sigma^*\Lambda}=\Gamma_{\sigma^*\Lambda}/\Gamma_{\sigma N}$ and $R_{\phi\Lambda}=\Gamma_{\phi\Lambda}/\Gamma_{\omega N}$. $R_{\sigma^*\Lambda}$ is obtained from the $\Lambda-\Lambda$ potential as $U_{\Lambda}^{\Lambda}(\rho_{{B}0})=-10$ MeV, which was extracted from the $\Lambda$ bonding energies of double-$\Lambda$ hypernuclei. $R_{\phi\Lambda}=-\sqrt{2}/2$ is corresponding to the SU(6) symmetry broken case~\cite{fortin17}. Here, the coupling between the $\Sigma$, $\Xi$ hyperons and $\sigma^*$ mesons are set as $R_{\sigma^*\Xi}=0,~R_{\sigma^*\Sigma}=0$, since the information about their interaction is absent until now. The values of $R_{\sigma Y}$ and $R_{\sigma^*\Lambda}$ with above constraints in different DDRMF effective interactions are shown in Table~\ref{table.HNcoup}. Here, for DD-MEX parameter series, we only select the DD-MEX set for investigation.
\begin{table}[htbp!]
	\centering
	\caption{The Coupling constants between hyperons and $\sigma$ meson, $\Gamma_{\sigma Y}$ and $\Lambda$-$\sigma^*$, $\Gamma_{\sigma^*\Lambda}$ in different DDRMF effective interactions.}\label{table.HNcoup}
	\scalebox{1}{
		\begin{tabular}{lcccccccccccc}
			\hline\hline
			&$R_{\sigma\Lambda}$ ~~&$R_{\sigma\Sigma}$  ~~&$R_{\sigma\Xi}$  ~~&$R_{\sigma^*\Lambda}$\\
			\hline 
			DD-LZ1       &0.610426 ~~&0.465708  ~~&0.302801  ~~&0.87595\\
			DD-MEX       &0.612811 ~~&0.469159  ~~&0.304011  ~~&0.86230\\
			DD-ME2       &0.609941 ~~&0.460706  ~~&0.302483  ~~&0.85758\\
			DD-ME1       &0.608602 ~~&0.457163  ~~&0.301777  ~~&0.85828\\
			DD2          &0.612743 ~~&0.466628  ~~&0.303937  ~~&0.86420\\
			PKDD         &0.610412 ~~&0.461807  ~~&0.302729  ~~&0.84965\\
			TW99         &0.612049 ~~&0.468796  ~~&0.303632  ~~&0.85818\\ 
			DDV          &0.607355 ~~&0.452777  ~~&0.301101  ~~&0.87979\\
			DDVT         &0.591179 ~~&0.399269  ~~&0.292391  ~~&0.92256\\
			DDVTD        &0.591108 ~~&0.399023  ~~&0.292352  ~~&0.92246\\
			\hline\hline
	\end{tabular}}
\end{table}

\subsection{Neutron Star from DDRMF model}\label{sec.NS}
\begin{figure}[htbp]
	\centering 
	\includegraphics[scale=0.5]{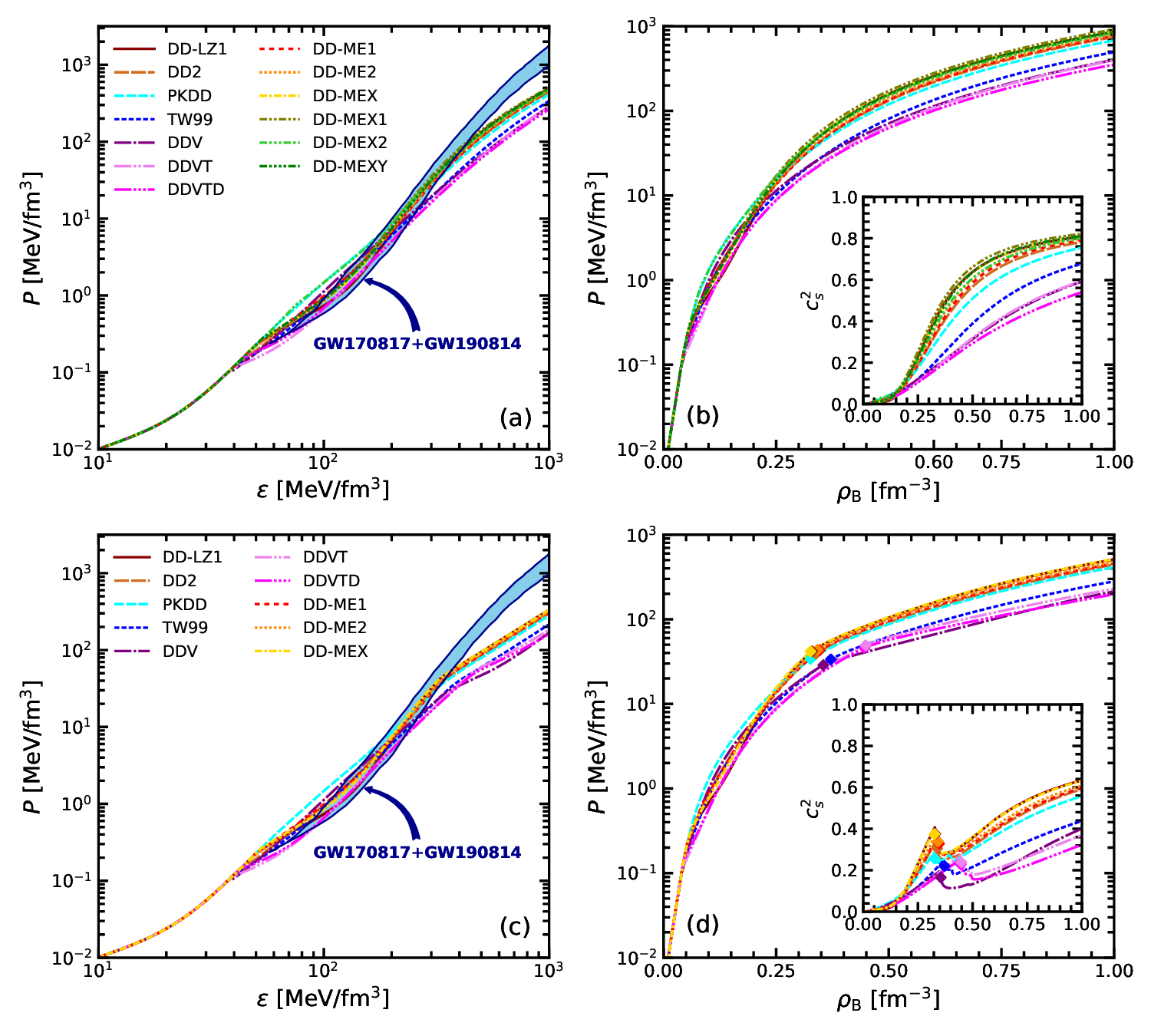}
	\caption{The pressure $P$ versus energy density $\varepsilon$ and the pressure $P$ versus baryon density $\rho_B$ of neutron star matter from DDRMF models. Panels (a) and panel (b) for the neutron star matter. Panels (c) and panel (d) for the hyperonic star matter. The joint constraints in panel (a) and panel (c) are from GW170817 and GW190814. The corresponding speeds of sound in units of the speed of light shown in subfigure of panel (b) and panel (c).}\label{fig.3}
\end{figure}   
Together with the conditions of beta equilibrium and charge neutrality in Eq. \eqref{eq.betacond} and Eq. \eqref{eq.chargecond}, {the EOSs of neutron star matter with DDRMF models can be obtained in panel (a) and panel (b) in Fig.~\ref{fig.3}, which shows the pressures of neutron star matter as a function of energy density and the pressures as functions of density, respectively}. The crust EOS of the non-uniform matter is generated by TM1 parameterization with Thomas-Fermi approximation ~\cite{bao15}. In the core of neutron star, EOSs of the uniform matter are calculated with various DDRMF sets discussed above. 
Their density-dependent behaviors are very similar with those in symmetric nuclear matter. At high density region, the stiffer group sets provide higher pressures due to the stronger vector potentials. The joint constraints on EOS extracted from the GW170817 and GW190814 are shown as a shaded band here. When the energy density is smaller than $600$ MeV/fm$^{3}$, the EOSs from stiffer group sets satisfy the constraints from the gravitational wave detection, while the pressures obtained from softer group sets start to be lower than the constraint band from $\varepsilon =300$ MeV/fm$^{3}$. 
The speeds of sound in neutron star matter, $c_s$ from softer group sets in the insert of panel (b) are around $0.6$ at $\rho_B=1.0$ fm$^{-3}$. They are much lower than those from stiffer group EOSs, which rapidly increase from $\rho_B=0.2$ fm$^{-3}$ and can reach around $0.8$ at high density. 

\begin{figure*}[!htbp]
	\centering 
	\includegraphics[scale=0.5]{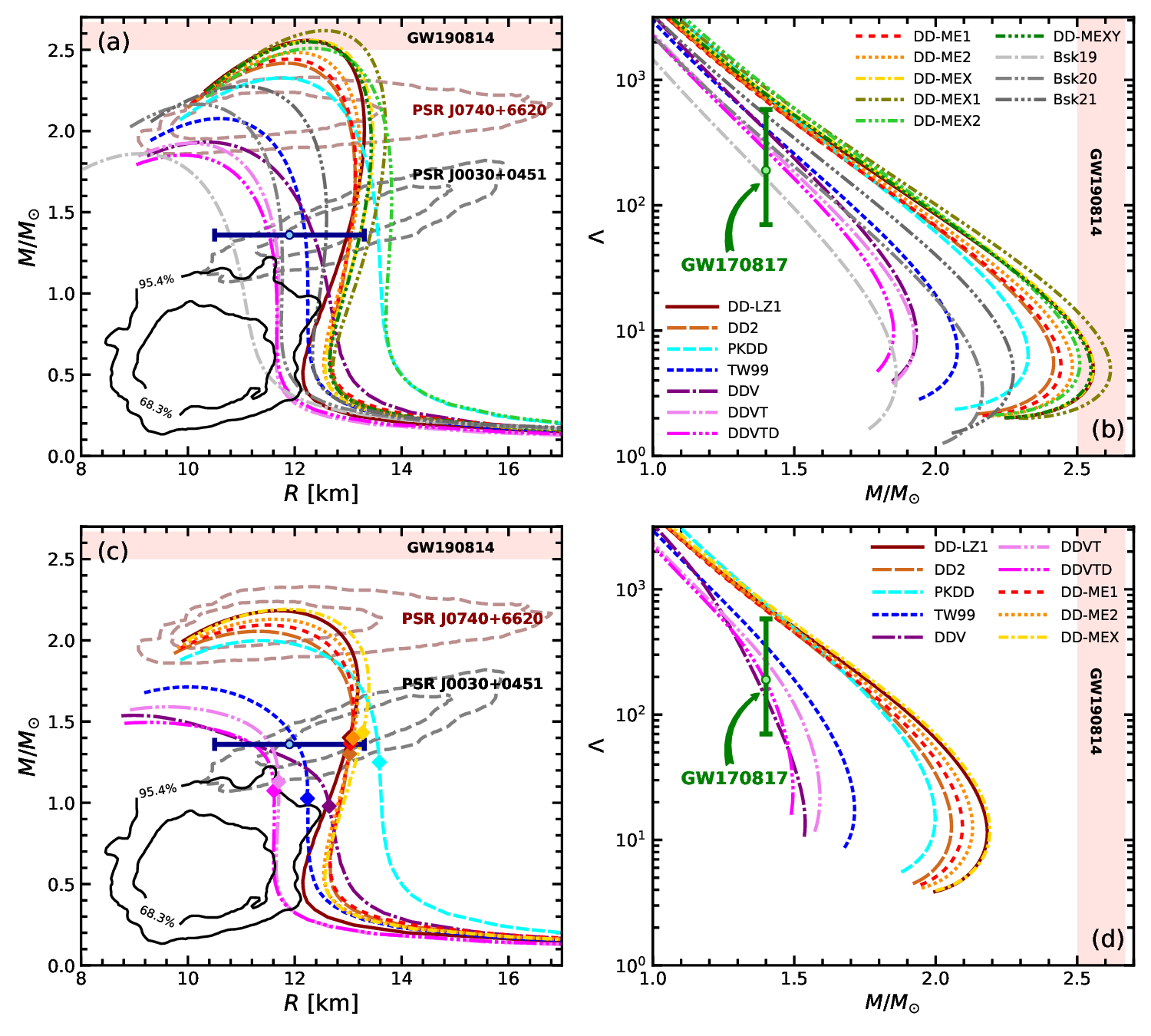}
	\caption{ Mass-radius relations and the tidal deformabilities as functions of star mass using the EOSs from different DDRMF sets. Panel (a) and panel (b) from the neutron star matter with different DDRMF models. Panels (c) and panel (d) from the hyperonic star matter. The dotted contours show the 68.3\% and 95.4\% credibility mass-radius constraints from PSR J0740+6620 \cite{miller21} and PSR J0030+0451 \cite{miller19}. The solid contours represent the central compact objects within HESS J1731-347 \cite{doroshenko22}. The horizontal error bar at 1.4$M_{\odot}$ is from GW170817 \cite{abbott18}. The horizontal error bar represents the tidal deformability constraint with a  range of $70<\Lambda_{1.4}<580$ from GW170817.}\label{fig.4}
\end{figure*} 
The mass-radius ($M-R$) relation of a static neutron star can be obtained by solving TOV equation in Eq. \eqref{eq.tov} with the EOS of neutron star matter as input. The $M-R$ relations from various DDRMF models are shown { in panel (a) and panel (b) of Fig.~\ref{fig.4}}. The $M-R$ relations from BSk series \cite{goriely10,potekhin13} are also shown for comparison. Additionally, we include mass-radius observations from measurements of PSR J0740+6620 and PSR J0030+0451 by NICER, the mass-radius constraints on the compact central object (CCO) from HESS J1731-347 \cite{doroshenko22}, as well as the radius constraint, $R_{1.4}=11.9\pm1.4$ km, from gravitational wave event GW170817 \cite{abbott18}. 
As depicted in the figure, nearly all these DDRMF parameter sets yield mass-radius relations that satisfy the $95.4\%$ confidence constraints of PSR J0740+6620 and PSR J0030+0451 while only the parameter sets in the stiffer group can satisfy the $68.3\%$ confidence constraints of PSR J0740+6620. The maximum masses calculated from the EOSs of the stiffer group are all above $2.3~M_\odot$, where DD-LZ1, DD-MEX, DD-MEX1, DD-MEX2, and DD-MEXY sets can even support the neutron star above $2.5-2.6~M_\odot$ because of their strongest repulsive contributions from $\omega$ meson, which are in accord with the observed mass of the secondary compact object in GW190814, $2.50-2.67~M_\odot$. Among these parameter sets, PKDD and DD-MEX2 produces the largest radius of $1.4 M_\odot$ due to its highest values of $L$ from Table \eqref{table.sat}, which exceeds the radius constraint from GW170817. 
What's more, these DDRMF parameter sets from the stiffer group fail to describe the measurement data from HESS J1731-347, while the EOSs from TW99, DDVT, DDVTD in the softer group, as well as the BSk series, { which have larger effective masses,} are capable of generating smaller radii in the lower mass region and can satisfy the $95.4\%$ confidence constraint from HESS J1731-347. In particular, the $M-R$ relation from BSk19 with smaller $L$ aligns with the $68.3\%$ uncertainties of HESS J1731-347 compared to BSk20 and BSk21.  
The GW170817 event also provides a valuable constraint on the tidal deformability, which corresponds to a radius of a $1.4M_{\odot}$ neutron star within the range of $70<\Lambda_{1.4}<580$ \cite{abbott18}, which is represented in panel (b) of Fig~\ref{fig.4} by a horizontal error bar, favoring the soft EOSs from the softer group of the DDRMF parameter sets, as well as the EOSs from BSk19, BSk20, and BSk21 sets.  

\begin{table}[htbp!] 
	\centering
	\caption{ Parameter $\Gamma_{\rho  N}(\rho_0)$ and $a_{\rho}$ generated from the DDVT model for different $L$ at the saturation point from $E_{\rm sym}$ fixed at $\rho_ B=0.11~\rm fm^{-3}$.} \label{table.differL}
	\begin{tabular}{r|ccccc}
		\hline \hline  
		$L$ [MeV]  &~~~ $\Gamma_{\rho N}(\rho_0)$ ~~~~~&$a_{\rho}$ \\
		\hline  
		26 & 7.250170  & 0.759445  \\
		30 & 7.367444  & 0.702782  \\
		40 & 7.637540  & 0.575940  \\
		42.35 & 7.697112  & 0.548702  \\ 
		\hline \hline
	\end{tabular}
\end{table} 
Next we want to further construct an EOS which can produce the $M-R$ relation that complies with the HESS J1731-347 observation constraints using the DDRMF model. We choose the DDVT set to calculate the core EOS since the DDVT set, which include the tensor coupling and can produce rather smaller radii at the low-mass region { due to its larger effective mass} from Fig. \eqref{fig.4} (a). 
The crust EOSs were derived from the IUFSU model at various values of $L$, with the symmetry energy fixed at $\rho_N=0.11~\rm fm^{-3}$ using the self-consistent Thomas-Fermi approximation \cite{bao14} and this choice based on the similarity of their nuclear saturation properties, which are listed in the last line in Table \eqref{table.sat}. We choose two extreme crust EOSs generated by the IUFSU family of models with $L=47,~110$ MeV, while the core EOSs will be obtained from the current family of DDVT parameterizations with $L=26,~30,~40$ MeV because a smaller $L$ for the core EOS can produce a smaller radius in the intermediate mass region of neutron stars \cite{huang24}. By manipulating the coupling strength of the isovector meson in the DDVT parameter set, the hadronic EOSs with $L=26,~30,~40$ MeV can been obtained. It should be noted that when the slope of the symmetry energy falls below $L=26$ MeV, the speed of sound in nuclear matter becomes less than zero. Table \ref{table.differL} displays the $\rho$ meson coupling constants for different $L$ and the couplings of the original DDVT model with $L=42.35$ MeV is also shown for comparison. It's obvious that there is a decrease in the $\rho$ meson coupling strengths at nuclear saturation density, $\Gamma_{\rho N}(\rho_0)$, as $L$ is reduced. 

\begin{figure}[htbp]
	\centering
	\includegraphics[scale=0.5]{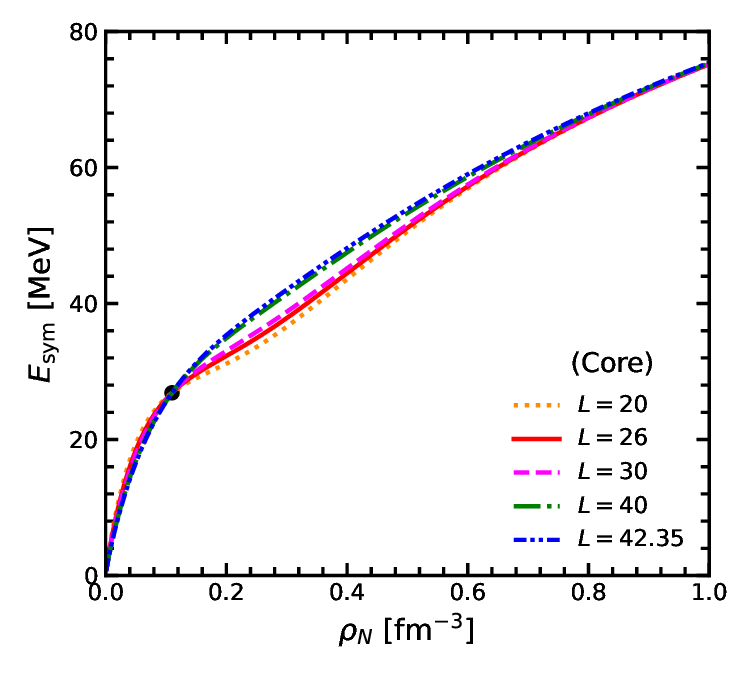}
	\caption{Symmetry energy $E_{\rm sym}$ as a function of the nucleon density for the DDVT sets with different $L$. The behaviors of the slope of symmetry energy $L$ are shown in the insert.} \label{fig.DDVT_differL_esym}
\end{figure}
The density dependence of the symmetry energy $E_{\rm sym}$ is plotted in Fig. \ref{fig.DDVT_differL_esym} for the various $L$ parameter sets presented in Table \ref{table.differL}, which plays a crucial role in determining the properties of neutron stars. Smaller values of the $L$ parameter yield larger symmetry energy values below the sub-saturation density, while exhibiting smaller values in the high-density region. Unlike the behavior of $E_{\rm sym}$ obtained from nonlinear RMF parameter sets like the TM1 model in Ref. \cite{ji19,hu20}, the density-dependent model's $E_{\rm sym}$ converges above a density of 0.8 $\rm fm^{-3}$ due to the last term in Eq. \eqref{eq.esym} tends to approach zero exponentially and the symmetry energy at high densities is primarily determined by the contribution of the first term.  

\begin{figure}[htbp]
	\centering
	\includegraphics[scale=0.5]{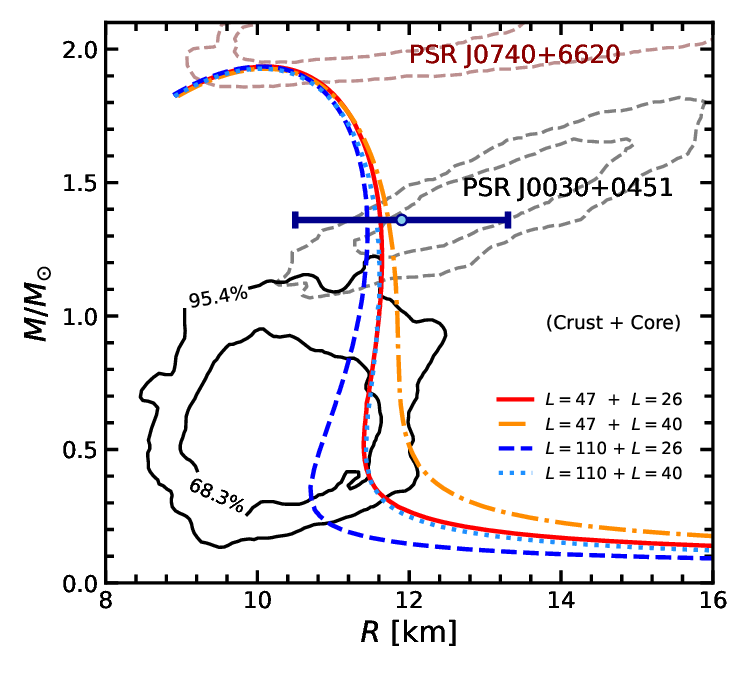}
	\caption{Mass-radius relations of neutron stars obtained using the sets from Table \eqref{table.differL}.} \label{fig.DDVT_differL_MR}
\end{figure} 
The corresponding $M-R$ relations for neutron stars with the constructed EOSs are shown in Fig. \ref{fig.DDVT_differL_MR}. It is observed that the $L$ parameter has minimal impact on the maximum mass and the corresponding radius of the neutron star.  The mass and radius of a CCO within the supernova remnant HESS J1731-347 have been estimated as $M=0.77^{+0.20}_{-0.17}M_{\odot}$ and $R=10.4^{+0.86}_{-0.78}$ km, respectively \cite{doroshenko22}, making the radius at $0.77M_{\odot}$ an important factor to consider. As $L$ increases from 47 MeV to 110 MeV, the radii at $0.77M_{\odot}$ have a reduction of approximately about 0.35 km, and as the $L$ of the core EOS decreases to $26-40$ MeV from the original value of $42.35$ MeV, the radii corresponding to the low-mass neutron stars obtained from the EOSs become sufficiently small to meet the $68.3\%$ credibility constraint, except for the combination of $L=47$ MeV for the crust EOS and $L=40$ MeV for the core EOS. Therefore, the observational data from HESS J1731-347 suggest the requirement of a crust EOS with a higher $L$ parameter and a core EOS with a lower $L$ parameter, representing an extremely soft EOS in both segments which is also consistent with the observables of PSR J0740+6620, PSR J0030+0451 from NICER, the GW170817 event, and the PREXII. This explains why the BSk19 EOS can also accurately describe HESS J1731-347. The properties of the neutron stars, i.e. maximum mass ($M_{\rm max}$), the corresponding radius ($R_{\rm max}$), the central density ($\rho_c$), the radius ($R_{\rm 1.4}$), and the dimensionless tidal deformability ($\Lambda_{\rm 1.4}$) at $1.4M_{\odot}$, as well as the radius ($R_{\rm 0.77}$) and the corresponding density at $0.77M_{\odot}$, can be seen in Ref. \cite{huang24}.

\subsection{Hyperonic Star from DDRMF Model}\label{sec.HS}    
Similarly, EOSs from DDRMF models for the hyperonic star matter are obtained in the { panel (c) and panel (d) of Fig.~\eqref{fig.3}}. The EOS of the inner crust is also chosen from the TM1 parameterization with self-consistent Thomas-Fermi approximation as before \cite{bao14} and the EOSs of the core region are calculated with various DDRMF parameter sets. They almost become softer from $\varepsilon\sim 300$ MeV/fm$^{3}$ compared to the neutron star matter in { Fig. \eqref{fig.3} (a)} due to the appearances of hyperons. In high-density region, all of them are below the joint constraints on the EOS from GW170817 and GW190814 events. The onset densities of the first hyperon are marked by the full discretized symbols, which are around $0.28-0.45$ fm$^{-3}$. $c^2_s$ of hyperonic star matter in { panel (d) of Fig. \eqref{fig.3}} is not smooth anymore since the appearance of hyperon can sharply reduce the magnitude of $c^2_s$, especially at first onset density. For the hard EoS, the $c^2_s$ becomes $0.6$ in hyperonic star matter from $0.8$ in neutron star matter at high-density region.

{ The $M-R$ relations of hyperonic star from  DDRMF parameter sets are shown in panel (c) of Fig.~\eqref{fig.4}}. The onset positions of the first hyperon in the relations are shown as the discretized symbols. After considering the strangeness degree of freedom, the maximum masses of the hyperonic star will reduce about $15\%\sim20\%$. Among these parameter sets, the DD-LZ1, DD-MEX, DD-ME2, DD-ME1, DD2, and PKDD sets generate the hyperonic star heavier than $2M_\odot$. Furthermore, the central densities of the hyperonic star become higher compared to the neutron star, all of which are above $5\rho_0$. The role of hyperons in the lower mass hyperonic star is strongly dependent on the threshold density of the first onset hyperon. The properties of a hyperonic star whose central density is below this threshold are identical to those of a neutron star. When the central density of the hyperonic star is larger than the threshold, the properties of hyperonic star will be influenced. { For the softer EOSs, the lower mass neutron stars are more easily affected. This is because the central densities at low-mass region of the neutron stars from the softer EOSs are much higher than the onset densities of the first hyperons. However, the central densities of the harder EOSs at low-mass region are almost lower than the onset densities of the first hyperons. For example, the central densities from the softer EOSs at $1.4M_\odot$ are about $0.50$-$0.58~\rm fm^{-3}$ and the onset densities of the first hyperons are about $0.35$-$0.45~\rm fm^{-3}$, so the appearance of hyperons has a big impact on the low mass region. But for the harder EOSs, the central densities at $1.4M_\odot$ are about $0.32$-$0.35~\rm fm^{-3}$ and the onset densities of the first hyperons are about $0.33$-$0.35~\rm fm^{-3}$, the appearance of hyperons have very little effect. Therefore, the radii of the hyperonic stars at $1.4M_\odot$ from DDV, DDVT, and DDVTD decrease from $12.3060,~11.6058,~ 11.4615$ km to $10.8990,~11.451 5,~10.9880$ km, a reduction of about $5\%$ compared to those of the neutron stars.}

The dimensionless tidal deformabilities of hyperonic star are plotted in { panel (d) of Fig.~\eqref{fig.4}}. For the harder EOSs, the hyperons only can influence the magnitudes of $\Lambda$ at the maximum star mass region, while they can reduce the dimensionless tidal deformability at $1.4M_\odot$ for the softer EOSs, such as TW99, DDV, DDVT, and DDVTD set, whose $\Lambda_{1.4}$ are closer to the constraints from GW170817.  Therefore, the compact stars in the GW170817 events may be the hyperonic stars.

The properties of neutron star and hyperonic star, i.e., the maximum mass ($M_{\rm max}/M_{\odot}$), the corresponding radius ($R_{\rm max}$), the central density ($\rho_c$), the radius ($R_{1.4}$) and dimensionless tidal deformability ($\Lambda_{1.4}$) at $1.4M_\odot$ from present DDRMF models can be found in our previous works \cite{huang20,huang22,huang24}. 

Now, the maximum masses of hyperonic stars are $1.50\sim 2.20 M_\odot$, and the corresponding radii are in the range of $9.30\sim 11.86$ km, which are reduced compared to the cases without considering the strangeness degree of freedom. The central densities become larger compared to those of neutron stars. The smallest radius of the hyperonic star at $1.4M_\odot$ is $10.90$ km from DDV, whose dimensionless tidal deformability is just $136$. In general, the maximum mass of a hyperonic star can exceed $2M_\odot$ if the EOS is a hard type, whose maximum mass approaches $2.2M_\odot$ with DDRMF model for a neutron star. Therefore, one solution to the ``hyperon puzzle" is to adopt the stiff EOS. The softer EOS only can describe the hyperonic star whose mass is around $1.5M_\odot$.

\section{Summary} \label{sec4}
The neutron star consisting of nucleons and leptons, and the hyperonic star considering additional contributions from hyperons were reviewed in the DDRMF model due to recent rapid achievements about astronomical observations on the compact star. The DDRMF parameter sets (DD-LZ1, DD-MEX, DD-MEX1, DD-MEX2, DD-MEXY, DD-ME2, DD-ME1, DD2, PKDD, TW99, DDV, DDVT, DDVTD) were adopted to calculate the properties of the neutron star and hyperonic star, which were created by reproducing the ground-state properties of several finite nuclei with different considerations.

The EOSs of symmetric nuclear matter and neutron star matter at high-density region are separated into the softer type (DDV, DDVT, DDVTD{ , TW99}) and stiffer one (DD-MEX, DD-MEX1, DD-MEX2, DD-MEXY, DD-ME2, DD-ME1, DD2, PKDD). The maximum masses of neutron stars generated by the softer EOSs cannot approach $2.0 M_\odot$, which can hardly satisfy the 68.3\% confidence constraints PSR J0740+6620. However, the radii of the corresponding neutron star are relatively small that can comply with the $95.4\%$ confidence constraint from HESS J1731-347 and their dimensionless tidal deformability at $1.4 M_\odot$ are around $275\sim510$ which are in accord with the value extracted from the GW170817 event.  Meanwhile, the harder EOS can lead to a very massive neutron star because of their strongest repulsive contributions from $\omega$ meson. In particular, the DD-MEX, DD-MEX1, DD-MEX2, DD-MEXY and DD-LZ1 parameter sets even can produce neutron stars with masses of  $2.5-2.6~M_\odot$, which can explain the secondary object in GW190814 with a mass of $2.50-2.67 ~M_\odot$.

To further study the light compact object in HESS J1731-347, a reasonable hadronic EOS has been obtained by manipulating the coupling strength of the isovector meson of the DDVT parameter set for the core EOS and the crust EOS is obtained by the IUFSU parameterization set \cite{bao14}. The observational data from HESS J1731-347 suggest the requirement of a crust EOS with a higher $L$ parameter and a core EOS with a lower $L$ parameter, representing an extremely soft EOS in both segments. The corresponding $M-R$ relations can also be consistent with the observables of PSR J0740+6620, PSR J0030+0451 from NICER, the GW170817 event. 

For the hyperonic stars, the meson-hyperon coupling strengths in DDRMF parameter sets were generated by the empirical hyperon-nucleon potential in symmetric nuclear matter at nuclear saturation density. The strangeness scalar and vector mesons were introduced to consider the $\Lambda-\Lambda$ potential in hyperonic star with the bond energies of double $\Lambda$ hypernuclei. The hyperonic star matter becomes softer compared to the neutron star matter. The onset densities of the first hyperon were around $2\rho_0\sim3\rho_0$ in present DDRMF models. The hyperon was raised earlier in the stiffer EOS and the appearance of hyperon can reduce the speed of the sound of the hyperonic star matter. The maximum mass of the hyperonic star is larger than $2M_\odot$ for the stiffer DDRMF parameter sets.  In addition, the hyperons influence the properties of the hyperonic star in the low-mass region from softer EOS since its central density is very higher. Therefore, dimensionless tidal deformability at $1.4 M_\odot$ will get smaller and be closer to the constraints from GW170817.

\section*{Acknowledgments} 
This work was supported in part by the National Natural Science Foundation of China (12175109).

\bibliography{Reference}

\end{document}